# Full-color nanorouter for high-resolution imaging

*Mingjie Chen[1,#], Long Wen[1,#], Dahui Pan[1], David R. S. Cumming[2], Xianguang Yang[1,\*] and Qin Chen[1,\*]*

[1]Institute of Nanophotonics, Jinan University, Guangzhou 511443, China

[2]James Watt School of Engineering, University of Glasgow, Glasgow G12 8QQ, UK

[*]chenqin2018@jnu.edu.cn, xianguang@jnu.edu.cn

[#]equal contribution

## Abstract

Pixel scaling effects have been a major issue for the development of high-resolution color image sensors due to the reduced photoelectric signal and the color crosstalk. Various structural color techniques have been proposed and demonstrated the large freedom in color manipulation by the structure design. However, the optical efficiency and the color distortion limit the practical applications due to its intrinsic filtering mechanism. Instead, on-chip full-color routing is quite charming for improving the signal-to-noise ratio. In this paper, a single-layer quick response code like nanorouter is proposed for the full-color light routing in a pixel level of image sensors. It shows much higher routing efficiency than various planar lens schemes for signal wavelength focusing. Moreover, over 60% signal enhancement with robust polarization insensitivity is obtained in all three primary color bands with a same nanorouter by a multi-objective optimization method. Negligible color distortion is observed from the reconstructed color image. Such a simple nanorouter structure is promising for the development of image sensor, photovoltaics and display.

# 1. Introduction

Since the film camera was replaced to the digital camera based on charge coupled device (CCD) or complementary metal-oxide-semiconductor (CMOS) image sensor (IS), it has been a main trend of pursuing small pixel sizes to meet the requirement of high-resolution digital imaging with several tens of millions of pixels [1, 2]. The state-of-the-art CMOS ISs have a pixel size entering a sub-micron scale [3]. The shrinkage of the pixel size of ISs raises serious signal-to-noise issues and brings challenges to conventional optical components [4] including dye color filters, microlens, etc. For example, the incident light flux of each pixel drops to one third by reducing the pixel size from 1.4 μm×1.4 μm to 0.8 μm×0.8 μm. Backside illumination (BSI) model has been commonly used in commercial high-resolution CMOS ISs to improve the filling ratio of active region and reduce the color crosstalk [5]. Moreover, many attempts have been recently made to design structural color filters that applying various nano-optical effects, including extraordinary transmission (EOT) [6, 7], metallic nanoantennas [8], Fano resonance [9], Mie resonance [10], guided mode resonance (GMR) [11], and so on. Compared to the conventional dye color filters based on material absorption, structural color techniques realize spectral filtering via artificial micro-/nano-structures with the advantages of CMOS process compatibility, stability and suppressed spatial color crosstalk [12]. Although thorough investigation has been made to explore the fundamental physics [13], grow high-quality materials [14] and optimize the fabrication and integration methods [15] of structural color techniques, none could beat dye color filters in terms of light transmission efficiency (~90%) and color purity [16]. In addition, most structural color filters are based on periodic nanostructures and show an obvious dependence of the filtering performance on the period number [6, 10, 12, 17]. Serious degeneration involving both optical efficiency and color purity occurs with the reducing period number for fitting in a single pixel [10]. For example, the peak transmission in red (R), green (G) and blue (R) bands is approximately 30% for 1.2 μm-squared size metallic nanohole filters [6]. Therefore, it is important to find out the intrinsic limitation of the optical efficiency in current ISs and put forward a new technique route to address this issue.

In current color imaging or display systems, illumination light is usually divided into several parts in each unit cell according to the spatial distribution of pixels, for example, four pixels in a Bayer array with a *RGGB* unit cell [18] or three pixels in a *RGB* array with a *RGB* unit cell [19]. In each pixel, light in one color band transmits through a color filter and is detected by a photodiode or human eyes.

Obviously, in such a configuration most light is filtered out without any contribution to the photoelectric or photobiological signal. For example, approximately one third light component is detected but two thirds are wasted in a *RGB* unit cell as shown in Figure 1(a), i.e., a maximum optical efficiency is only 33% even if the color filter has transmittance as high as 100%. Alternatively, routing light to appropriate directions determined by its wavelength rather than excluding the unwanted light components by the filters is expected to provide higher optical efficiencies [20, 21]. As shown in Figure 1(b), the incident light flux of a unit cell is routed to *R*, *G* and *B* pixels according to the wavelengths respectively. Theoretically, this spatial dispersion allows all photons with different energy to be detected in appropriate photodiodes with a maximum optical efficiency of 100%. Traditionally, diffractive gratings are widely used to direct multiwavelength light to different spatial positions. However, its bulky size limits its applications in ISs. Plasmonic antennas have shown the color sorting functions in a subwavelength scale based on the wavelength-dependent excitation of surface plasmon resonance and the asymmetrical interference [19, 22-25]. Even a single silver nanorod was found to be able to scatter light into several beams with different colors based on its longitudinal multipolar plasmon modes [26]. However, the intrinsic high loss of metallic nanostructures limits the overall optical efficiency for imaging applications. In contrast, dielectric metasurfaces allow remarkable optical manipulation with extremely low loss [27-31]. A full-color router was reported with a GaN metalens, where multiwavelength routing was realized by integrating 4×4 spatial multiplex nanopillars into a complex unit cell [21]. However, the router has a dimension of 50 μm × 50 μm with a focal length of 110 μm, which is not appropriate for the integration in ISs. Moreover, the routing efficiencies of *R*, *G* and *B* colors drop to 15-38% as a cost of multiplexing. Scattering by dielectric nanoantennas has also been explored for color routing [20,32]. Panasonic Corporation demonstrated a color nanosplitter integrated CCD with a pixel size of 1.43 μm×1.43 μm for color imaging, where a SiN nanostrip deflects polarized light to different pixels according to the wavelength [32]. The total amount of light received by four pixels in a 2×2 unit cell was found to be 1.71 times that of the color filter method using a Bayer array [16]. NTT Corporation used a dual-nanoantenna structure to reduce the polarization dependence of color routing and achieved an improvement of the total detected power in a unit cell of 2.90 times compared to the color filter method [16] when inserting such nanorouters on a glass slide between a target image and a monochrome IS [20]. However, these two methods suffer from large spectral crosstalk because only a part of pixels are covered with antennas, i.e.

the overall improvement of the detected light power mainly comes from the nonselective light transmission in those white (W) pixels. For example, the normalized light transmission of the *G* pixel is less than 1 without any enhancement [20].

In this paper, we proposed quick response (QR) code like full-color nanorouters of high optical efficiency and high color purity for the applications in ISs with a micron-scale pixel size. Such single-layer color nanorouters consist of titanium oxide nanoblocks in a QR-code like array in a low refractive index environment on the surface of a monochrome BSI IS, which is easy to fabricate and compatible to the CMOS processes. The QR-code like configuration was optimized by the Non-dominated Sorting Genetic algorithms-II (NSGA-II). Full-color routing with low polarization dependence was demonstrated in a 1.1 μm×1.1 μm-pixel IS with an improvement of 2.88 times of the total detected power in a unit cell, where an average optical efficiency of each pixel is enhanced by a factor of 1.66 with a color quality of 0.63. Moreover, 30% improvement on optical efficiency together with a color quality as good as the dye color filters are also demonstrated. The reconstructed image from a standard multispectral image shows excellent color fidelity and obvious improvement of brightness.

## 2. Design method and monochrome nanorouters

When designing nanophotonic structures including structure color filters [6-12] and color nanorouters [20 21, 32], a particular geometry like nanoholes and gratings is ordinarily selected from qualitative considerations, and then parameter sweeps of dimensions and materials are employed. Such an intuitive design based on the fundamental optical principles and the brute-force search approach probably misses the best solution and confronts extremely low efficiency for hyper-dimensional and multi-objective optimization due to the huge amount of computing resources. Recently, a gradient-based optimization approach was used in the color nanorouter design and showed excellent wavelength-dependent spatial light sorting [33-35]. However, a two-dimensional (2D) model cannot reveal the actual physics and directly instruct the practical device design [35]. Both a design element of 10 nm×10 nm and the 3D porous structures in a nanoscale are far beyond the fabrication capability of current standard IS processes [34, 35]. Instead, here QR-code like nanorouters are proposed and optimized by an algorithm of NSGA-II, which has advantages in multi-objective optimization [36, 37]. As shown in Figure 1c, such single-layer color nanorouters consist of high refractive index nanoblocks in a QR-code like array in a low refractive index environment. In a case of transmissive filtering configuration, multiplex pixels can increase the flux ratio of a specific color. Therefore, the Bayer array with two green pixels in each unit cell (2×2 pixels) is commonly used in ISs because human eyes are sensitive to green light. However, in a case of perfect color routing configuration, extra green pixel won't increase the ratio of green color. So, a *RGB* unit cell as shown in Figure 1b is adopted to reflect the fundamental property of nanorouters for the application of ISs. Nanoblocks distribute inside each pixel in an area of 1.1×1.1 μm$^2$, where each block has a lateral dimension of 100 nm×100 nm for a feasible design considering current fabrication capability. The block distribution is different in *R*, *G* and *B* pixels but the distribution in each unit cell repeats over the whole pixel array [38, 39].

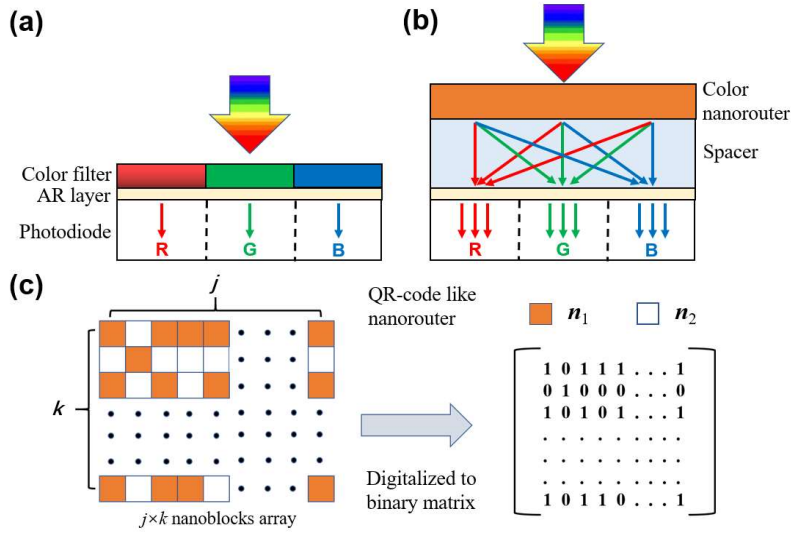

Figure 1. Schematic of spectral engineering in an IS. (a) An IS with color filters. (b) An IS with color nanorouters. (c) Schematic of a QR-code like color nanorouter.

To optimize the QR-code like distribution of nanoblocks, the genetic algorithm (GA) is applied, which is a global optimization technique inspired by the biological heredity and evolution. GA utilizes a set of binary strings to characterize features of the objectives. During the iteration, the outstanding designs of previous generation are selected based on the fitness and served as parents to generate the next generation. The binary strings of parents would be split and randomly recombined, which is similar to the chromosome division and recombination. Mutation is introduced to increase diversity in order to escape from the local optimal solutions by flipping the value of binary strings. The performance of offspring is evaluated according to the fitness calculated by the defined fitness function. These processes would be repeated until some objectives are met. Following the logic of GA, such a QR-code like distribution of the nanoblocks can be digitalized to a binary *j*×*k* matrix as shown in Figure 1c, where '1' represents the high refractive index nanoblock region and '0' represents the low refractive index environment. Finite-difference time-domain algorithm is applied to solve Maxwell's Equations. Light transmission to silicon diode of each pixel is served as the fitness of the binary structure. To achieve the goal of maximizing transmission for each pixel, NSGA-II as a multi-objective optimized version of GA is applied, where a set of non-dominated solutions called Pareto front reach a good trade-off between objectives for each generation. Based on the above method, nanorouters for *G* color were first designed, where the nanorouters cover three pixels and are expected to sort the green light incident on the whole unit cell to the middle pixel. As shown in Figure 2a, each pixel is divided into a 11×11 binary matrix. $TiO_2$ is chosen as the high refractive index material of nanoblocks. It has been widely used to construct various meta-optics devices in the visible region due to its transparency and relatively high refractive index [40-42]. The $TiO_2$ nanoblock layer has a thickness of 600 nm to provide sufficient phase difference for spatial dispersion. Such a high aspect ratio $TiO_2$ post array has been successively demonstrated in metalens [40, 41]. A $SiO_2$ spacer is sandwiched between the nanoblocks and the silicon diodes.

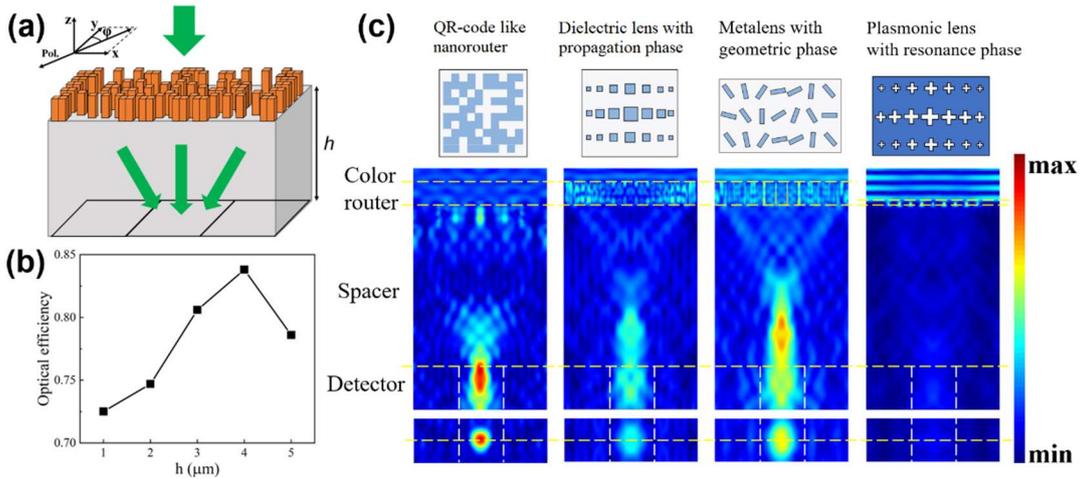

Figure 2. (a) Schematic of a nanorouter for *G* color in an IS with a *RGB* unit cell. (b) Optical efficiencies at different spacer thicknesses at a wavelength of 550 nm with *y*-polarized normal incidence. (c) Cross-section electric field distributions of a QR-code like nanorouter, dielectric lens with propagation phase, metalens with geometric phase and plasmonic lens with resonant phase. The thickness of spacer *h* is set to 4 μm. Yellow lines indicate the positions of nanostructures, spacers and detectors. White dashed lines indicate the position of each pixel.

Optical efficiency of the nanorouter for sorting green light to the middle pixel is defined as the ratio of the collected flux of the middle pixel to the incident flux over the unit cell (three pixels) at a wavelength of 550 nm. For the conventional dye filter scheme, the maximum efficiency is 30% assuming 90% transmittance of the dye filter. In contrast, the nanorouter are able to collect the incident light on the neighboring pixels to the middle pixel as an antenna resulting in an improved optical efficiency. As shown in Figure 2b, the improvement is quite robust to the spacer thickness with an enhancement factor between 2.4-2.8 times compared to the color filtering scheme based on the dye filters. The light routing phenomenon is clearly seen from the electric field distribution in Figure 3. At the surface of the silicon diode, the electric field is concentrated inside the middle pixel as expected with weak crosstalk to the other two pixels. The relatively uniform electric field at the top surface of nanorouters gradually concentrates to the middle pixel in both vertical and lateral directions. This promising result demonstrates a remarkable improvement of optical efficiency of the color routing scheme over the color filtering scheme, which is a key point in the further development of super-resolution ISs. It is interesting to compare the optimization results of the proposed structure to the ones based on the fundamental optical principles such as dielectric planar lens with propagation phase [41], metalens with geometric phase [40] and plasmonic flat lens with resonant phase [43]. All these planar lens designs have the same lateral size of 3.3 μm×1.1 μm as the nanorouters. The lenses with the propagation phase and the geometric phase have the same thickness 600 nm as the nanorouters but the resonant phase lens has a thickness of 50 nm due to its high absorption loss. As seen from Figure 2c, all lenses show light focusing but the electric field intensity at the focuses are much weaker than the nanorouters. It is attributed to the limited numerical aperture (NA) due to the small lateral size of the lenses [44]. The coarse phase distribution and sparse light channels limited by the feature size of the nanostructures in a fixed NA also reduces the interference at the right focus. In contrast, the GA algorithm generated design shows a more robust light routing.

### 3. Full-color routing and imaging

For full-color imaging in practical applications, a multi-wavelength router is required to enhance the optical efficiency in each color band. Two approaches including sectoring and interleaving are usually used to design multi-wavelength or multifunctional metasurfaces [21, 33]. Generally, devices are subdivided into regions that are individually designed for a single wavelength, where the sectored device regions are stitched together at a scale larger than a wavelength and the interleaved device regions are mixed in a subwavelength scale [33]. For example, full-color routing at visible light was demonstrated in a GaN metalens with the interleaved configuration [21]. The optical efficiencies drop from 50%-92% to 15-38% due to the crosstalk between different interleaved regions although the lateral size is as large as 50 μm. As shown in Figure 2c, the performance of the lenses designed with the optical principles is significantly limited, which is more difficult for a multi-wavelength router. The algorithm assisted design including the gradient inverse method and the GA method has a large design freedom to balance the choices of multiple dimensional parameters and intrinsically involve all the potential physical effects including the coupling between nanostructures. Therefore, it will be more efficient for a multi-objective optimization.

The optical efficiencies of the designed nanorouters are plotted in Figure 3a, where all three color bands show better performance than the average efficiency of current dye filtering technique indicated by the dashed line. The peak efficiencies of *R*, *G* and *B* pixels are all above 50% with an enhancement factor in a range of 1.63-1.81 times. It means that the actual light flux of a specific color in each pixel of such a nanorouter integrated IS is nearly the same as a conventional IS with a 1.63-1.81 times larger pixel size. The improvement of the signal-to-noise ratio encourages the further development of the current IS technique to sub-micron regions. Figure 3b shows the electric field distributions at the surface of silicon diodes at 425 nm, 550 nm and 675 nm, where light at three wavelengths concentrates to the expected pixels respectively. The results demonstrate the feasibility of the color nanorouters for efficient color sorting even for a multi-band requirement. Although the degeneration of optical efficiency is observed similar to the previously mentioned sectoring and interleaving metasurfaces due to the crosstalk between multiplexed design [20, 21], the efficiency as high as 50% is still remarkable considering the current color filtering technique. An effective light flux enhancement factor is defined as the ratio of the collected light flux in the target color band to that of the dye filter scheme. As shown in Figure 3d, all the color filter schemes have small enhancement factors (<1). In particular, both EOT and GMR schemes have an average enhancement factor less than 0.4, where the EOT one suffers from its low transmission and the GMR one suffers from its narrowband resonance although it has high transmission. In contrast, all the color routing schemes show larger enhancement factors above 1. Although the EOT structure [45,46] and the multilayer stack [47] have been integrated in ISs and shown the expected color filtering function, the color routing schemes is more promising in terms of the optical efficiency. The best results are from the gradient based inverse design [34] with an enhancement factor of 2.67, where the feature size of 10 nm×10 nm and 3D randomly porous profile in a nanoscale require more effort in fabrication techniques. The proposed QR-cord like design is simple and has an average enhancement factor of 1.66, which is four times larger than the EOT scheme.

Apart from optical efficiency, color quality is another important performance specification. Color quality is evaluated by the differences of hue (H), saturation (S) and value (V) of the colors, indicating the difference between the detected color and the indexed color [10, 48]. The color quality becomes zero when the detected color is close to a different primary color and it is equal to one if the detected color perfectly matches the indexed color. Actually, the color quality has been a bottleneck issue for the structural color technique especially in a transmissive configuration [6-11]. For example, the transmission and the linewidth are contractionary in metallic nanohole color filters based on the EOT effect [6]. The improvement of optical efficiency at a cost of large crosstalk is not expected. Generally, the dye filters based on the material absorption demonstrate sharp absorption edges to suppress the spectral crosstalk [16]. Using such dye filters, the optical efficiencies of each pixel in a *RGB* unit cell is shown as dashed lines in Figure 3c. Assuming the *R*, *G* and *B* bands are 600nm-700nm, 500nm-600nm and 400nm-500nm, the spectral crosstalk is very small. As shown in Figure 3d, the average color quality of the *R*, *G* and *B* bands of the dye color filter scheme is as high as 0.88. In contrast, plasmonic color filtering techniques such as EOT phenomenon [6] usually have small average color qualities due to the broad resonance linewidth limited by the high absorption loss. Mie scattering filters also suffer from the low color qualities due to its scattering property [10]. Both two have an average color quality around 0.5. Metallic guided mode resonance (GMR) filters have less absorption loss due to its field concentration in the dielectric waveguide and thus provide a narrowband resonance resulting in an improved color quality of 0.69 [11]. A large color quality of 0.79 can be obtained using multilayer stacks but the different stack structures of the *R*, *G* and *B* bands increase the complexity of fabrication [47]. In the single-pixel color deflector schemes demonstrated by Panasonic [32] and NTT [20], the designed *W+B* and *W+R* pixels inevitably induce flat profiles of the detected power spectra, which greatly increase the spectral crosstalk. For example, the NTT scheme has a color quality less than 0.4. The gradient based inverse design shows a large color quality of 0.75 due to the small limitation of the structure profile. The QR-code like scheme has a decent color quality of 0.63. There is one way to further reduce the spectral crosstalk of the QR-code like design for some applications with high requirements on the color quality. The dye color filters can be integrated underneath the nanorouters to filter out the crosstalk in each pixel and improve the color quality. In this case, the nanorouters can be integrated onto the current ISs without major modification of the standard processes. As shown in Figure 3c, the combined design with both dye color filters and nanorouters demonstrates three distinct spectra in *R*, *G* and *B* bands with small spectral crosstalk. At the same time, the optical efficiencies overcome the conventional dye color filters in all three bands. As seen from Figure 3d, in the combined structure the color quality increases to 0.9 with an average enhancement factor of 1.33.

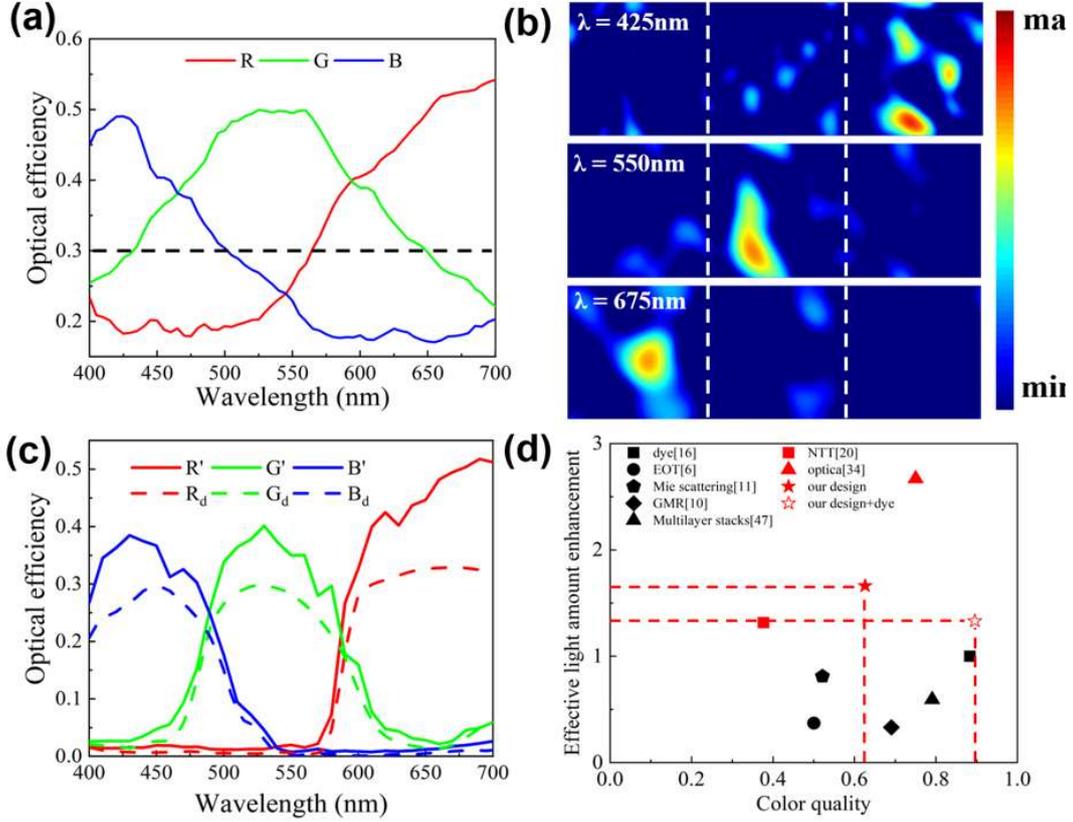

Figure 3. (a) Optical efficiencies of the nanorouter in *R*, *G* and *B* bands. Dashed line shows the average efficiency of current dye filtering technique. (b) Electric field distributions at the surface of silicon diodes at 425 nm, 550 nm and 675 nm. The spacer thickness is 4 μm. White dashed lines indicate the position of each pixel. (c) Optical efficiencies of the scheme combining nanorouters and dye filters. Dashed lines show the efficiencies of the scheme with only dye filters. (d) Comparison of the average color quality and the average effective light flux enhancement factor between various filters and routers with different mechanisms.

Except the polarization imaging application [49], the optoelectronic responses of ISs are expected to have no polarization dependence. The dye color filters based on material absorption properties naturally satisfy this requirement. However, the structural schemes including structural color filters and nanorouters may show strong polarization dependence due to the asymmetric design [32, 34, 50]. A *RGB* unit cell as shown in Figure 1b is intrinsically asymmetric. To address this issue, each design of the nanorouters is simulated twice under orthogonal polarized incidence respectively. The average transmittance obtained from these two simulations is taken as the fitness in the GA optimization to ensure low polarization sensitivity of the optimal nanorouters. Optical efficiencies of the nanorouter for various polarization angles are shown in Figure 4a-c. It is seen that all the spectra show quite robust profiles for a polarization angle in a range of 0-180°. The spectral correlation coefficients for three color bands shown in Figure 4d keep above 0.9 in all cases indicating the polarization insensitivity of the nanorouter.

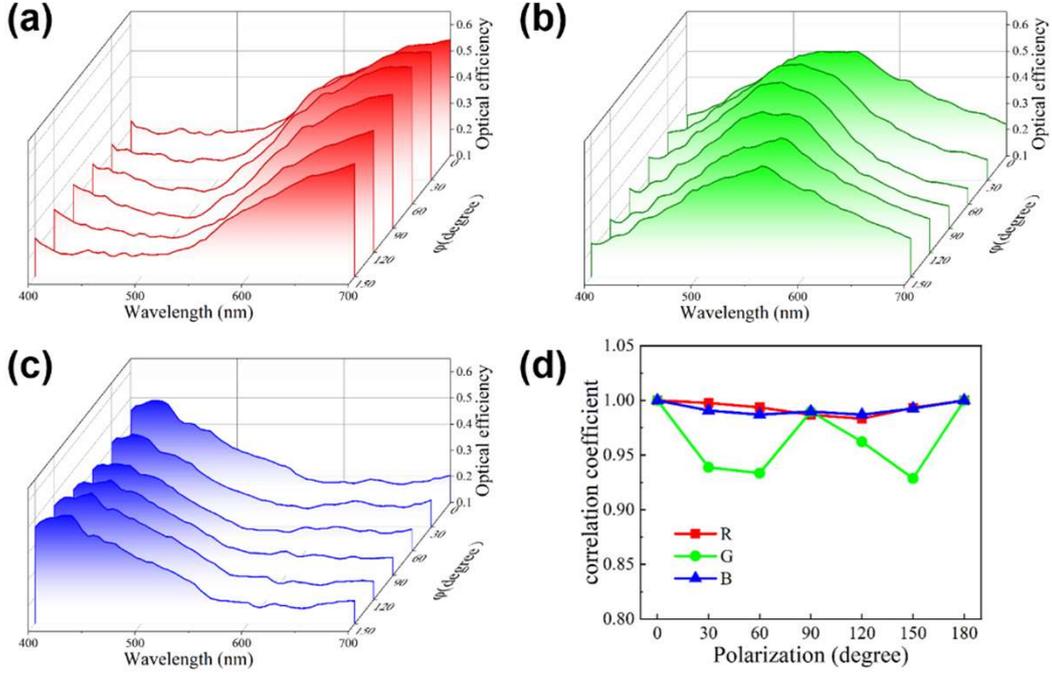

Figure 4. (a)-(c) Optical efficiencies of the nanorouter in *R*, *G* and *B* bands for various polarization angles. (d) Spectral correlation coefficients for various polarization angles. The nanorouter is the same as Figure 3b.

Finally, it is interesting to predict the imaging performance with the full-color nanorouters. Although there is no experimental results, the actual imaging process can be modeled with a multispectral target image based on a conversion matrix method [20] as shown in Figure 5a. Because the spatial spectral information *L* of the target image is known, the detected photocurrent vector *X* can be calculated by the product of *L* and the spectral responses *S* of the dye filters or the nanorouters. In digital color imaging, the *RGB* signal vector *Y* is converted from the detected photocurrent vector *X* of *R*, *G* and *B* pixels by a 3×3 conversion matrix *A*, i.e. *Y* = *AX*. Three rectangular spectra *L'* in 400nm-500nm, 500nm-600nm and 600nm-700nm are defined to obtain *X'*, which is applied to solve *A* with the associated *RGB* signal vector *Y'*. A multispectral image is selected from an open access website [24] as shown in Figure 5a. The reconstructed images with the *RGB* value *Y* with dye filters and nanorouters are shown in Figure 5b and c respectively. Both images exhibit an excellent color quality with the original multispectral image. To visualize the difference in signal intensities of two schemes, the spectral integration of the collected flux in each pixel is linearly converted to gray values. The results of both dye filters and nanorouters are shown in Figure 5d and e respectively. It is obvious that the nanorouters scheme shows much larger signal intensity. Therefore, such a nanorouter integrated IS scheme with high optical efficiency and good color quality is promising for high resolution imaging application.

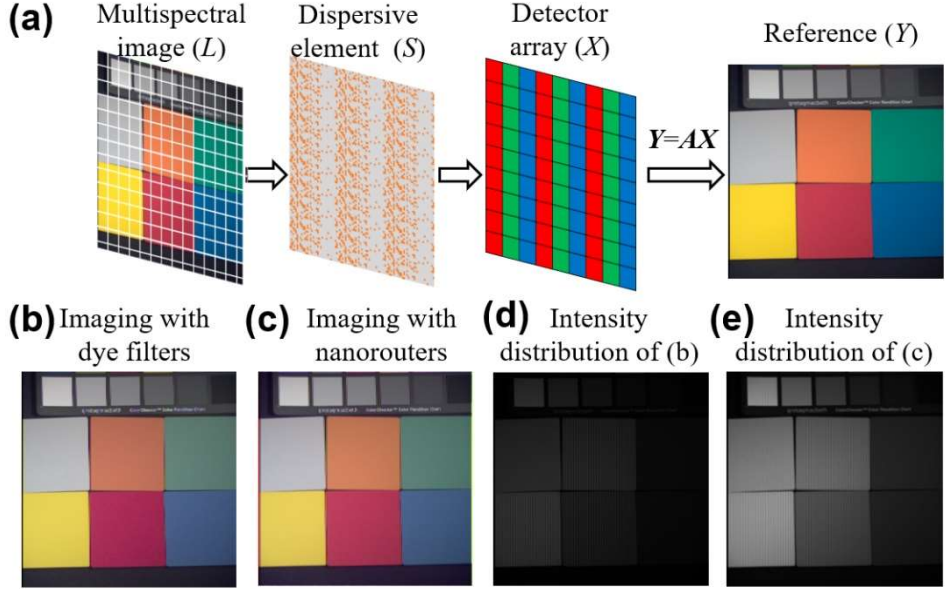

Figure 5. (a) Simulated imaging processes based on a multispectral image. (b) Reconstructed color image with dye filters ignoring the brightness. (c) Reconstructed color image with nanorouters in Fig. 3a ignoring the brightness. (d) and (e) are the grey-scale images of (b) and (c), respectively.

## 4. Conclusion

A QR-code like nanorouter is proposed for full-color routing in IS. Compared to various structure color techniques and the recently reported light routing techniques, the QR-code like nanorouter enables high optical efficiency with negligible color distortion in a simple structure, which is important for the development of low-cost ISs for high-resolution applications. The pixel level light routing function at multiple wavelengths overcomes various plasmonic and metasurface lenses. Over 60% signal enhancement is demonstrated in a 1.1 μm×1.1 μm pixel level compared to the state-of-the-art dye filter scheme. The proposed method is promising not only in ISs but also display and photovoltaics.

**Acknowledgements**

We are grateful for financial supports from National Key Research and Development Program of China (No. 2019YFB2203402), National Natural Science Foundation of China (Nos. 11774383, 92050108, 11774099, 11874029 and 11804120), Guangdong Science and Technology Program International Cooperation Program (2018A050506039), Guangdong Basic and Applied Basic Research Foundation (No. 2020B1515020037), Pearl River Talent Plan Program of Guangdong (No. 2019QN01X120). DRSC is supported by UK EPSRC Grant EP/T00097X/1.

**Conflict of Interest**

The authors have declared no conflict of interest.

Data available on request from the authors